\documentclass[12pt]{article}

\newsavebox{\ns}
\newsavebox{\dbrane}
\newsavebox{\dbshort}

\usepackage{epsfig}
\usepackage{amssymb}

\def\be{\begin{eqnarray}}
\def\ee{\end{eqnarray}}

\newcommand{\nn}{\nonumber}
\newcommand\para{\paragraph{}}
\newcommand{\ft}[2]{{\textstyle\frac{#1}{#2}}}
\newcommand{\eqn}[1]{(\ref{#1})}

\def\Dslash{\,\,{\raise.15ex\hbox{/}\mkern-12mu D}}
\def\Dbarslash{\,\,{\raise.15ex\hbox{/}\mkern-12mu {\bar D}}}
\def\delslash{\,\,{\raise.15ex\hbox{/}\mkern-9mu \partial}}
\def\delbarslash{\,\,{\raise.15ex\hbox{/}\mkern-9mu {\bar\partial}}}
\def\pslash{\,\,{\raise.15ex\hbox{/}\mkern-9mu p}}
\def\calDslash{\,\,{\raise.15ex\hbox{/}\mkern-12mu {\cal D}}}

\begin{document}
\pagestyle{plain}
\setcounter{page}{1}
\newcounter{bean}
\baselineskip16pt

\begin{titlepage}

\begin{center}
\today
\hfill hep-th/0307302\\
\hfill MIT-CTP-3403 \\

\vskip 1.5 cm
{\large \bf Monopoles in the Higgs Phase}
\vskip 1 cm 
{David Tong}\\
\vskip 1cm
{\sl Center for Theoretical Physics, 
Massachusetts Institute of Technology, \\ Cambridge, MA 02139, U.S.A.\\} 
{\tt dtong@mit.edu}

\end{center}

\vskip 0.5 cm
\begin{abstract}
We describe new solutions of Yang-Mills-Higgs theories consisting of magnetic 
monopoles in a phase with fully broken gauge symmetry. Rather than spreading out radially, 
the magnetic field lines form flux tubes. The solution is topologically stable 
and, when embedded in ${\cal N}=2$ SQCD, preserves $1/4$ of the supercharges. 
From the perspective of the flux-tube the monopole appears as a kink. 
Many monopoles may be threaded onto a single flux tube and placed 
at arbitrary separation to create a stable, BPS necklace of solitons. 

\end{abstract}

\end{titlepage}


Should we ever be lucky enough to find a magnetic monopole, one might 
consider displaying it in the Natural History Museum embedded 
within a superconductor. The magnetic flux lines would not spread out 
radially, but instead have the peculiar property of forming a flux tube. 
Adjoining interactive displays could describe this 
delightful consequence of the Meissner effect while waxing lyrical about 
an analogous mechanism in QCD which is responsible for holding us all together. 
Nearby, the holographic image of a celebrity physicist might explain how similar 
strings are conjectured to underlie the very fabric of our universe.
\para
In this paper we shall describe  smooth, topologically stable, magnetic monopole 
solutions with the property described above. 
Recall that in QED, monopoles are Diracesque singular affairs, essentially put 
into the theory by hand. To find smooth solutions, we must turn to $SU(2)$ 
Yang-Mills theories. When the gauge group is broken to $U(1)$ by an adjoint 
scalar field, 't Hooft and Polyakov showed that topological considerations  
guarantee the existence of monopoles \cite{tp}. However the theory is in the 
Coulomb phase and the magnetic flux lines spread  
out radially. Suppose we attempt to naively break the gauge symmetry further 
so that the $U(1)$ is also Higgsed at low-energies. The magnetic field lines must now form 
flux tubes at large distances, but the price we have paid is to lose the topological 
stability of the configuration which remains, at best, meta-stable. An exception 
to this is if $U(1)\rightarrow Z_2$ which can be achieved by a second adjoint scalar field. 
In this case $Z_2$ strings are supported and the resulting stable 
monopole-flux tube configuration was discussed by 
Hindmarsh and Kibble \cite{hk}. Monopoles attached to $Z_N$ strings have 
also been discussed in \cite{marco}. 
\para
Here we shall discuss a slightly different symmetry breaking structure, 
involving a locking of gauge and flavour symmetries, which 
supports both $U(1)$ flux tubes of the familiar Nielsen-Olesen form \cite{no} 
and magnetic monopoles \`a la 't Hooft-Polyakov \cite{tp}. We work with an  
${\cal N}=2$ supersymmetric theory in $d=3+1$ dimensions with a $U(N)_G$ vector 
multiplet and $N_f=N$ fundamental hypermultiplets\footnote{This is a minimal choice: 
the solutions we describe exist for any $N_f\geq N$.} with an  
$SU(N)_F$ flavour symmetry. The full symmetry group is\footnote{The classical theory 
has a further $SU(2)_R\times U(1)_R$ R-symmetry group, but 
this will not be responsible for stabilising any soliton solutions and we shall 
pay it less attention.},
\be
G=U(N)_G\times SU(N)_F
\nn\ee
The bosonic field content of the theory 
is as follows: the vector multiplet contains a $U(N)_G$ gauge field $A_\mu$, 
together with a complex adjoint scalar field $\phi$; the hypermultiplets 
contain scalars $q_i$, $i=1,\ldots, N_F$, each of which transforms in 
the ${\bf N}$ representation of $U(N)_G$, and a further $N_f$ scalars 
$\tilde{q}_i$ transforming in the $\bar{\bf N}$. The bosonic part of the Lagrangian 
is given by,
\be
{\cal L}&=&{\rm Tr}\left(\frac{1}{4e^2}F_{\mu\nu}F^{\mu\nu}
+\frac{1}{2e^2}|{\cal D}_\mu\phi|^2\right)
+ \sum_{i=1}^{N_f}\left(|{\cal D}_\mu q_i|^2+|{\cal D}_\mu\tilde{q}_i|^2\right)
\nn\\&& 
 - {\rm Tr}\left(\frac{1}{2e^2}[\phi^\dagger,\phi]^2+
e^2|\sum_{i=1}^{N_f}q_i\tilde{q}_i|^2+ \frac{e^2}{2}
(\sum_{i=1}^{N_f}q_iq_i^\dagger-\tilde{q}_i^\dagger\tilde{q}_i-v^2)^2\right) 
\nn\\&&
- \sum_{i=1}^{N_f}\left(q^\dagger_i|\phi-m_i|^2q_i+\tilde{q}_i|\phi-m_i|^2
\tilde{q}_i^\dagger\right)
\nn\ee
In the above expression we have introduced complex mass parameters $m_i$ and a 
real FI parameter $v^2$, each consistent with ${\cal N}=2$ supersymmetry. 
For generic values of these parameters the theory has a unique vacuum state, up 
to Weyl permutations, given by,
\be
\phi={\rm diag}(m_i)\ \ \ \ ,\ \ \ \ \ q^a_{\ i}=v\delta^a_{\ i}\ \ \ \ \ ,\ \ \ \ \ 
\tilde{q}^a_{\ i}=0
\label{vac}\ee
where $a=1,\ldots,N$ is the colour index. The $U(N)_G$ gauge symmetry is completely 
broken and the theory lies in a gapped, 
colour-flavour-locked phased. 
\para
The pattern of symmetry breaking at intermediate energy scales 
depends on the relative values of $m_i$ and $v^2$. For $|m_i-m_j| \gg ev$, the flavour 
group is explicitly broken by the masses at a higher scale than the spontaneous 
symmetry breaking induced by the FI parameter,
\be
U(N)_G\times SU(N)_F
\stackrel{m}{\longrightarrow} U(1)^N_G\times U(1)_F^{N-1}
\stackrel{v}{\longrightarrow}U(1)^{N-1}_{\rm diag}
\label{break1}\ee
However, if $ev\gg |m_i-m_j|$, then the spontaneous breaking due to the vacuum 
expectation value of $q$ occurs at a higher scale than the explicit breaking due 
to masses, 
\be
U(N)_G\times SU(N)_F\stackrel{v}{\longrightarrow} SU(N)_{\rm diag}
\stackrel{m}{\longrightarrow} U(1)^{N-1}_{\rm diag}
\label{break2}\ee
For both patterns \eqn{break1} and \eqn{break2} 
the symmetry breaking due to the masses supports magnetic monopoles 
($\Pi_2(SU(N)/U(1)^{N-1})={\bf Z}^{N-1}$) while the symmetry breaking due to the 
FI parameter breaks a $U(1)$ factor, ensuring the stability of 
vortices ($\Pi_1(U(1))={\bf Z}$). Moreover, the full symmetry breaking enjoys the 
topology required to support both 
monopoles and fluxes. We shall now see that indeed the 
theory admits magnetic monopoles attached to two vortex strings which whisk 
away their flux. 
\para
\newcommand{\onefigurenocap}[1]{\begin{figure}[h]
         \begin{center}\leavevmode\epsfbox{#1.eps}\end{center}
         \end{figure}}
\newcommand{\onefigure}[2]{\begin{figure}[htbp]
         \begin{center}\leavevmode\epsfbox{#1.eps}\end{center}
         \caption{\small #2\label{#1}}
         \end{figure}}
\begin{figure}[htbp]
\begin{center}
\epsfxsize=4.5in\leavevmode\epsfbox{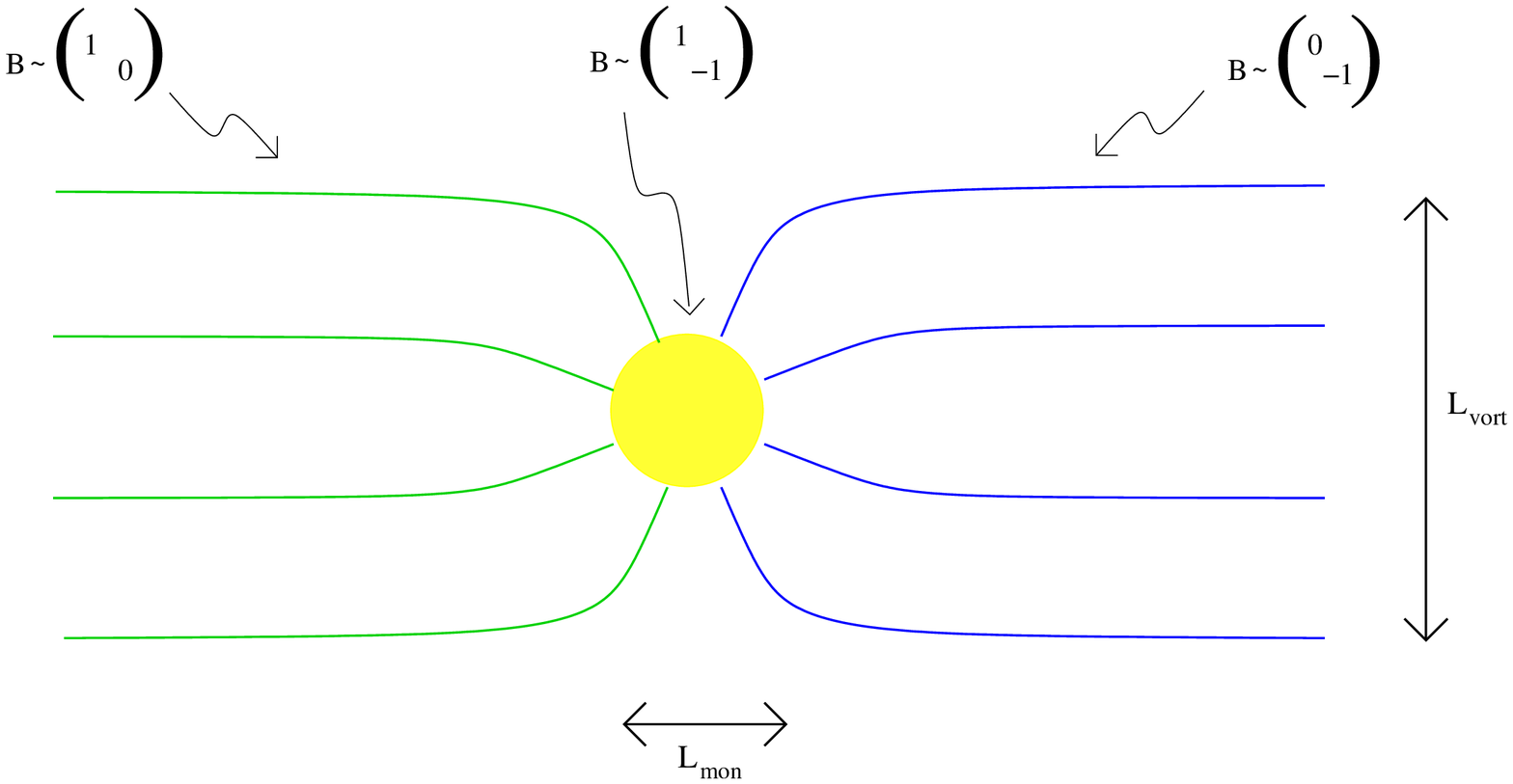}
\end{center}
 {\small An impressionistic rendering of the $U(2)$ monopole in the Higgs phase 
when $L_{\rm vort}\gg L_{\rm mon}$.}
\end{figure}
The solutions will turn out not to involve the fields $\tilde{q}$ and  
we set them to zero at this stage. Moreover, the simplest configurations 
have ${\rm Im}(m_i)=0$ which allows us to also set ${\rm Im}(\phi)=0$. 
In the following $\phi$ will therefore denote a real adjoint scalar field\footnote{It seems 
likely that interesting dyonic monopole-flux tube configurations can 
be built by relaxing this condition to allow ${\rm Im}(m_i)\neq 0$.}. Since 
the flux will leave the monopole in a tube, we must decide in which direction 
this string will head: we choose the $x^3$ direction.
Restricting to time independent configurations 
the Hamiltonian reads,
\be
{\cal H}&=& \frac{1}{2e^2}B_\rho^2+\frac{1}{2e^2}|{\cal D}_\rho\phi|^2 
+|{\cal D}_\rho q_i|^2+\frac{e^2}{2}(q_iq^\dagger_i-v^2)^2+q^\dagger_i(\phi-m_i)^2q_i
\nn\\ &=&
\frac{1}{2e^2}({\cal D}_1\phi-B_1)^2 + \frac{1}{2e^2}({\cal D}_2\phi-B_2)^2 
+({\cal D}_3\phi-B_3-e^2(q_iq_i^\dagger-v^2))^2 \nn\\ && 
+|{\cal D}_1q_i-i{\cal D}_2q_i|^2+|{\cal D}_3q_i+(\phi-m_i)q_i|^2 
-v^2B_3+\frac{1}{e^2}\partial_\rho(\phi B_\rho) \nn\\ &\geq&
-v^2B_3+\frac{1}{e^2}\partial_\rho(\phi B_\rho)
\label{ham}\ee
where we have left colour indices and traces implicit, summed over the flavour index 
$i$, and introduced the spatial index $\rho=1,2,3$. Both terms in the final line are 
topological invariants. The first measures the flux carried by vortex strings lying 
in the $x^3$ direction; the second measures the magnetic charge carried by a monopole. 
As we shall see, we can have strings without any need for monopoles, but the 
presence of a monopole will require two, semi-infinite vortex 
strings to carry away its 
flux.  In the Coulomb phase, the integral of $\partial\cdot (\phi B)$ is evaluated 
on the ${\bf S}^2_\infty$ boundary. In the present case the 
monopole's flux does not make it to all points on the boundary and is instead captured 
by integrals over the two planes ${\bf R}_{\infty}^2$ at $x^3=\pm\infty$. 
The Bogomoln'yi equations can be found within the  
total squares on the second line of \eqn{ham} and read
\be
B_1={\cal D}_1\phi\ \ \ ,\ \ \ B_2={\cal D}_2\phi\ \ \ ,\ \ \ 
B_3={\cal D}_3\phi+e^2(\sum_{i=1}^Nq_iq_i^\dagger -v^2)\nn\\
{\cal D}_1q_i=i{\cal D}_2 q_i\ \ \ ,\ \ \ 
{\cal D}_3 q_i=-(\phi-m_i)q_i\hspace{2.5cm}
\label{bog}\ee
A quick glance reveals these to be interesting mix of the 
monopole and vortex equations. I have not been able to find an explicit solution. 
Indeed, given that no analytic solution exists to the Nielsen-Olesen vortex equations, 
it seems rather unlikely that the task is any simpler for these generalised equations.
Nevertheless, 
we can gain insight into the form of the solution by studying the equation 
in two different limits.
\para
Let us start by considering the limit $|m_i-m_j|\gg ev$. 
The equations in the second line of \eqn{bog} 
can be solved simply 
by $q_i=0$, while, if we ignore the effect of $v^2$ for now, 
the equations in the top line become the familiar 
Bogomoln'yi equations $B_\rho\approx{\cal D}_\rho\phi$ describing a monopole 
with a non-abelian core of width $L_{\rm mon}\sim 1/|m_i-m_j|$. For 
distances $L \geq L_{\rm mon}$, the magnetic field lies primarily within the 
Cartan subalgebra $U(1)^{N-1}_G\subset SU(N)_G\subset U(N)_G$ and 
emerges radially from the monopole core. However, this radial 
behaviour cannot continue indefinitely. At scales 
$L_{vort}\sim 1/ev \gg L_{\rm mon}$, the effect of the Higgs mechanism 
becomes apparent, damping the magnetic field 
as can be seen from the third of the Bogomoln'yi equations in \eqn{bog}. At this 
point, it becomes energetically favourable to set the scalar fields $\phi$ 
and $q_i$ to their vacuum expectation values \eqn{vac} in order to allow 
the magnetic field $B$ to vanish throughout the bulk. However, the magnetic flux from the 
monopole has to go somewhere. To see where, note that when $\phi$ is set 
to its constant expectation value and $A_\rho$ is restricted to lie in the 
Cartan subalgebra then the non-trivial equations of \eqn{bog} read
\be
B_3=e^2(\sum_{i=1}^Nq_iq^\dagger_i-v^2)\ \ \ ,\ \ \ \ {\cal D}_1q_i=i{\cal D}_2q_i
\label{vort}\ee
which are the non-Abelian form of the familiar Abelian vortex equations. 
They describe a tubes of magnetic flux of width $L_{\rm vort}$ 
lying in the $x^3$ direction. 
The string has finite tension $2\pi v^2$, and therefore infinite mass due 
to its infinite length. This reflects the fact that, like quarks in QCD, 
monopoles in the Higgs phase do not like to be alone. 
\para
For a Dirac monopole in QED, the 
flux string is expected to depart in only one direction. When this 
happens, the tension of the string causes the monopole to accelerate 
and the configuration is unstable. However, for the superconductor example of 
the opening paragraph this situation is avoided as the Cooper pair 
condensate has charge 2 which allows for the formation of strings 
carrying a half quantum of  flux \cite{vol}. 
Thus the flux from the monopole may be carried 
away by two flux tubes of equal tension, leaving in opposite directions. 
Here we shall see that the solution to \eqn{bog} has a similar 
property where each flux tube now carries a single quantum of flux lying in 
a different $U(1)\subset U(N)_G$ subgroup. To see this, we turn to the 
opposite limit $ev \gg |m_i-m_j|$ where the width of the vortex $L_{\rm vort}$ 
is much smaller than the width of the monopole core $L_{\rm mon}$. There 
is now no spatial region in which the monopole looks like the usual 't Hooft-Polyakov 
radial configuration. However, we can make progress by studying the monopole 
from the perspective of the vortex string.
In fact, let us start by considering 
the situation $m_i=0$, so that the symmetry breaking is 
simply $G\rightarrow SU(N)_{\rm diag}$. The theory now supports vortex 
strings, but not monopoles. The vortices satisfy the equations \eqn{vort} and 
were studied recently in \cite{meami} (related systems were examined even 
more recently in \cite{yung}). For a 
single vortex of unit winding number (${\rm Tr}\,\int d^2x\, B_3=-2\pi$), it was 
shown that the surviving $SU(N)_{\rm diag}$ group acts on the soliton  
resulting in a moduli space ${\cal V}_N$ of solutions, 
\be
{\cal V}_N\cong {\bf C}\times {\bf CP}^{N-1}
\nn\ee
where ${\bf C}$ parameterises the center of mass of the vortex string 
in the $x^1-x^2$ plane, while ${\bf CP}^{N-1}$ describes the internal 
degrees of freedom of the vortex arising from the $SU(N)_{\rm diag}$ action. 
The K\"ahler class of ${\bf CP}^{N-1}$ is $2\pi/e^2$ \cite{meami}.
The low-energy dynamics of the vortex string can be described by a $d=1+1$ 
dimensional sigma model with target space ${\cal V}_N$. Since the vortex 
is BPS \cite{jose}, the low-energy dynamics preserves ${\cal N}=(2,2)$ supersymmetry.
\para
How is this picture changed by the introduction of masses $m_i$? The masses 
break the $SU(N)_{\rm diag}$ symmetry in the pattern \eqn{break2}, lifting the 
${\bf CP}^{N-1}$ moduli space. For a vortex of unit winding number there are 
now $N$ isolated solutions corresponding to an abelian 
vortex embedded diagonally in one of the $U(1)\subset U(N)_G$ subgroups. These 
different solutions are related by discrete $SU(N)_{\rm diag}$ transformations but, 
as this includes the action of a global symmetry group, are physically 
distinguishable configurations. From the perspective of the low-energy dynamics, 
the masses $m_i$ can be thought of as inducing a potential $V$ on ${\bf CP}^{N-1}$ 
with $N$ isolated minima. 
In fact, the exact form of the potential $V$ can be determined using the 
techniques described in \cite{me} and is of the form $V\sim k^2$, where 
$k$ is a Killing vector on ${\bf CP}^{N-1}$. The existence of multiple 
isolated vacua in the vortex theory gives rise to the possibility of a new object: 
a kink. Kinks (a.k.a. domain walls) in supersymmetric 
theories have received a lot of attention in the literature, starting with 
\cite{mirjam,qkink}. We shall now show that the kink on 
the flux tube is identified with the magnetic monopole.
\para
Rather than enter into the details of the $U(N)_G$ theory, here we simply 
concentrate on the case of $U(2)_G$ gauge group for which the internal vortex moduli 
space is ${\bf CP}^1$. We parameterise ${\bf CP}^1$ by a circle fibration over an 
interval, with $\psi\in [0,2\pi)$ labeling the circle, and 
$-\pi/e^2 \leq r\leq \pi/e^2$ labeling the interval. The circle degenerates 
at $r=\pm \pi/e^2$ to yield the topology of the sphere. 
Writing the two mass parameters as $(m_1,m_2)=(m,-m)$, the 
low-energy internal dynamics of the vortex string is governed by a 
$d=1+1$ dimensional massive sigma-model with ${\bf CP}^1$ target space,
\be
{\cal L}_{\rm vort}=\ft12 H(r)(\partial r)^2+\ft12 H^{-1}(r) (\partial \psi)^2 
- 2m^2 H^{-1}(r)
\nn\ee
where 
\be
H(r)=\frac{1}{\pi/e^2+r}+\frac{1}{\pi/e^2-r}
\nn\ee
The kinetic terms are those of a sigma-model on ${\bf CP}^1$ endowed with the 
round metric, while the potential 
term is proportional to the length${}^2$ of the $\partial_\psi$ Killing vector on 
${\bf CP}^1$. 
As we described above, the masses have lifted the moduli space of 
vortices, leaving behind two isolated configurations at the minima of the 
potential $r=\pm\pi/e^2$. These correspond to vortices carrying magnetic 
flux $B_3\sim{\rm diag}(0,1)$ and $B_3\sim{\rm diag}(1,0)$ respectively. 
\para
We now consider the kink on the flux tube interpolating between these 
two vacua. Such a string starts at $r=-\pi/e^2$ at $x^3\rightarrow -\infty$ 
and concludes at $r=+\pi/e^2$ as $x^3\rightarrow+\infty$. In fact 
domain walls in massive ${\bf CP}^N$ sigma-models of this type have been 
much studied in the literature, starting in \cite{qkink}. The solution is simply
\be
r=\frac{\pi}{e^2}\tanh(m(x^3-x_0))\ \ \ ,\ \ \ \psi={\rm const.}
\nn\ee
where $x_0$ is the center of mass of the kink along the string. 
From the perspective of the $d=3+1$ gauge theory, this kink on the vortex 
worldsheet is simply the monopole described by equations \eqn{bog}. 
To see this, firstly note that the 
mass of the domain wall is $4\pi m/e^2$, in agreement with the 
mass of the monopole calculated from the final term in \eqn{ham}. 
Secondly, we can 
examine the fluxes carried by the vortex string. As $x^3\rightarrow -\infty$, 
the $U(2)$ magnetic field lies in $B_3\sim{\rm diag} (1,0)$, while for  
$x^3\rightarrow +\infty$, the magnetic field lies in $B_3\sim {\rm diag}(0,1)$. Taking 
into account the direction of the flux, we see that the domain wall acts as 
a magnetic source of the form $B\sim {\rm diag}(1,-1)$. This is  
precisely the flux emitted by the monopole.  Note that the vortex 
preserves half the original ${\cal N}=2$ supersymmetry \cite{jose} and 
the domain wall preserves half the supersymmetry of the vortex theory 
\cite{mirjam,qkink}. The monopole-flux-tube-combo 
is therefore a $1/4$-BPS state in ${\cal N}=2$ SQCD. An impressionistic, 
and not entirely accurate, portrait of the magnetic flux lines is offered in the figure. 
\para
It is interesting to note that the original fascination with domain walls 
in ${\bf CP}^1$ sigma-models derived from the observation that they exhibit 
features reminiscent of magnetic monopoles \cite{qkink}. 
Here we provide a simple explanation for this fact: the domain walls {\it are} 
magnetic monopoles. 
The monopoles in question lie in the Higgs phase, and are therefore restricted to 
sit on a string of flux wherein they appear as domain walls. 
\para
For $U(N)_G$ gauge group, the situation is similar. There are now $N$ vacua 
of the low-energy vortex dynamics, and one can consider domain walls 
interpolating from the first vacuum ($B\sim{\rm diag}(1,0,\ldots,0)$) to the last 
($B\sim{\rm diag}(0,\ldots,0,1)$). Such domain walls were 
studied in detail in \cite{gtt}. It was shown that the kinks can be 
placed at arbitrary separation without experiencing attractive or repulsive 
forces. From the 
perspective of monopoles, this corresponds to the fact that (so called 
$(1,1,\ldots,1)$) monopoles can be threaded on a flux tube and placed 
arbitrary separation. They may slide along the string at will and are 
constrained only in that they may not pass each other. This results in a 
BPS necklace of monopoles, acting like hard beads threaded on a vortex flux tube. 
\para
Let us close by recalling two other areas of physics where solutions similar 
to those discussed above appear. 
The first sits on a tabletop: the A-phase 
of superfluid ${}^3$He supports configurations analogous to a monopole 
emitting one (or more) vortex strings \cite{vol}. This composite 
object is referred to as a nexus. In the case of ${}^3$He, the strings are 
supported by a global symmetry but similar configurations with gauged 
vortices are argued to appear in chiral p-wave superconductors \cite{vol}. 
The second application is in the context of cosmology. Configurations of the type 
discussed here have been invoked as a way to 
catalyse monopole-anti-monopole annihilation. This could be of 
interest either in the early universe to rid us of GUT monopoles 
\cite{lp}, or in the current epoch where necklaces of monopoles 
have been suggested as a source for ultra-high-energy cosmic 
rays \cite{bv}. It is to be hoped that the existence of the 
simple Bogomoln'yi equations \eqn{bog} may be of help in determining 
the dynamics of solitonic necklaces.

\subsection*{The Acknowledgments}
My thanks to Joyce Berggren, Neil Constable, Kimyeong Lee, 
Paul Townsend, Jan Troost and Grisha Volovik for useful conversations and 
comments and also 
to Sue Moran for driving while the Bogomoln'yi equations were derived. This work 
was undertaken at the Benasque Center for Science during the 
workshop on String Theory and I would like to thank the organisers and participants 
for providing such a stimulating and pleasant environment. 
I'm supported by a Pappalardo fellowship and am grateful to   
the Pappalardo family for their generosity. This work was also supported in part 
by funds provided by the U.S. Department of Energy (D.O.E.) under 
cooperative research agreement \#DF-FC02-94ER40818.

\end{document}